\documentclass[jcph]{apjrnl}

\usepackage{epsf}

\begin{document}

\title{Stellar Structure Modeling using a Parallel Genetic Algorithm for Objective Global Optimization}

\author{Travis S. Metcalfe$^{\ref{tac}}$ and Paul Charbonneau$^{\ref{hao}}$ \\
\additem[cfa]{Harvard-Smithsonian Center for Astrophysics, 60 Garden Street, 
Cambridge, MA 02138, USA},
\additem[tac]{Theoretical Astrophysics Center, Institute of Physics and
Astronomy, Aarhus University, 8000 Aarhus C, Denmark},
\additem[hao]{High Altitude Observatory, National Center for Atmospheric 
Research, P.O. Box 3000, Boulder, CO 80307-3000, USA} \\
E-mail: tmetcalfe@cfa.harvard.edu, paulchar@hao.ucar.edu}

\date{22 November 2002}

\maketitle

\begin{abstract}  
Genetic algorithms are a class of heuristic search techniques that apply
basic evolutionary operators in a computational setting. We have designed
a fully parallel and distributed hardware/software implementation of the
generalized optimization subroutine {\tt PIKAIA}, which utilizes a genetic
algorithm to provide an objective determination of the globally optimal
parameters for a given model against an observational data set. We have
used this modeling tool in the context of white dwarf asteroseismology,
i.e., the art and science of extracting physical and structural
information about these stars from observations of their oscillation
frequencies. The efficient, parallel exploration of parameter-space made
possible by genetic-algorithm-based numerical optimization led us to a
number of interesting physical results: (1) resolution of a hitherto
puzzling discrepancy between stellar evolution models and prior
asteroseismic inferences of the surface helium layer mass for a DBV white
dwarf; (2) precise determination of the central oxygen mass fraction in a
white dwarf star; and (3) a preliminary estimate of the astrophysically
important but experimentally uncertain rate for the $^{12}{\rm C}
(\alpha,\gamma)^{16}{\rm O}$ nuclear reaction. These successes suggest
that a broad class of computationally-intensive modeling applications
could also benefit from this approach.
\end{abstract}

\begin{keywords}
Stellar Structure; Parallel Computation; Optimization
\end{keywords}

\section{\bf Astrophysical Context \label{INTROSEC}}

About 5 billion years from now, the hydrogen fuel in the center of the Sun
will begin to run out and the helium that has collected there will begin
to gravitationally contract, increasing the rate of hydrogen burning in a
shell surrounding the core. Our star will slowly bloat into a red
giant---eventually engulfing the inner planets, perhaps even the Earth. As
the helium core continues to contract under the influence of gravity, it
will eventually reach the temperatures and densities needed to fuse three
helium nuclei into a carbon nucleus (the $3\alpha$ reaction). Another
nuclear reaction will compete for the available helium nuclei at the same
temperature: the carbon can fuse with an additional helium nucleus to form
oxygen. The amount of oxygen produced during this process is largely
determined by the relative rates of these two competing reactions
\cite{ww93}. Since the Sun is not very massive by stellar standards, it
will never get hot enough in the center to produce nuclei much heavier
than carbon and oxygen. These elements will collect in the center of the
star, which will then shed most of its red giant envelope---creating a
planetary nebula---and emerge as a hot white dwarf star \cite{kd00}.

Once a white dwarf star forms and the nuclear reactions have ceased, its
structural and thermal evolution is dominated by cooling, and regulated by
the opacity of its thin atmospheric outer layers. It will slowly fade as
it radiates its residual thermal energy into space---eventually cooling
through a narrow range of temperatures that will cause it to vibrate in a
periodic manner, sending gravity-driven seismic waves deep through the
interior and bringing information to the surface in the form of brightness
variations. This is fortunate, because a detailed record of the nuclear
history of the star is locked inside, and pulsations provide the only
known key to revealing it.

We can determine the internal composition and structure of pulsating white
dwarfs using the techniques of high speed photometry to observe their
variations in brightness over time, and then matching these observations
with a computer model which behaves the same way. The observational
aspects of this procedure have been addressed by the development of the
Whole Earth Telescope (WET) network \cite{nat90}, a group of astronomers
at telescopes around the globe who cooperate to produce nearly continuous
time-series photometry of a single target for 1-2 weeks at a time. The
Fourier spectra of such observations reveal dozens of excited modes with
periods in the range 100--1000 seconds, supporting our interpretation of
them as non-radial oscillations with gravity as the restoring force
($g$-modes). The WET has now provided a wealth of seismological data on
the different varieties of pulsating white dwarf stars.

The physical property of white dwarf models that most directly determines
the pulsation frequencies is the radial profile of the Brunt-V\"ais\"al\"a
(buoyancy) frequency, which is given by
\begin{equation}
  N^2 = -g \left( {d \ln \rho \over dr} - 
  {1 \over \Gamma_1} {d \ln P \over dr} \right) ,
\end{equation}
where $g$ is the local gravity, $\rho(r)$ the density, $P(r)$ the
pressure, and $\Gamma_1$ is $(\partial\ln P / \partial\ln\rho)$ at
constant entropy. The magnitude of $N^2$ reflects the difference between
the actual and the adiabatic density gradients, and sets the local
propagation speed of internal gravity waves. The observed frequencies, in
turn, are a measure of the average (inverse) wave speed in the portion of
the interior where the waves propagate. Inferring the $N^2$ internal
profile from the observed pulsation frequencies is thus a classical {\it
inverse problem}, on par in scope and complexity with similar problems
encountered in helio- and geo-seismology.

Consider first the complementary {\it forward problem}, which consists in
computing the oscillation frequencies of a {\it given} white dwarf
structural model. The forward modeling procedure begins with a static,
non-rotating, unmagnetized, spherically symmetric model of a pre-white
dwarf, which we allow to evolve quasi-statically until it reaches the
desired surface temperature. The models must initially satisfy two of the
basic equations of stellar structure: the condition of hydrostatic
equilibrium, which balances the outward pressure gradient against the
inward pull of gravity
\begin{equation}
  {dP \over dr} = {G M_r \over r^2}\rho\ ,
\end{equation}
and the continuity equation ensuring mass conservation
\begin{equation}
  {dM_r \over dr} = 4 \pi r^2 \rho\ ,
\end{equation}
where $M_r$ is the mass contained within a spherical shell of radius $r$.
White dwarf stars are compact objects supported mainly by electron
degeneracy pressure ($P_e$), and we can describe the core with a simple
polytropic equation of state of the form
\begin{eqnarray}
  P_e \propto \left( {\rho \over \mu_e} \right)^{5/3} ,
\end{eqnarray}
where $\mu_e$ is the mean molecular weight per free electron. Cooling is
achieved by leaking the internal thermal energy through the opacity of the
thin atmospheric layers at a rate consistent with the star's luminosity,
and adjusting the interior structure accordingly. Although we initially
ignore a third equation of stellar structure (which ensures thermal
balance), we do use it to evolve the models in a self-consistent manner.
The cooling tracks of our polytropic models quickly forget the unphysical
initial conditions and converge with the evolutionary tracks of
self-consistent pre-white dwarf models well above the temperatures at
which the hydrogen- and helium-atmosphere white dwarfs are observed to be
pulsationally unstable \cite{woo90}.

Next, the $g$-mode pulsation frequencies ($\sigma_g$) of the evolved
models must be calculated for comparison with the observations. Working in
the usual spherical polar coordinates $(r,\theta,\phi)$, the first step is
to express the radial displacement ($\Xi_r$) experienced by an oscillating
fluid element as
\begin{equation}
\Xi_r(r,\theta,\phi,t) =\xi_r(r) Y_\ell^m(\theta,\phi) \exp(i\sigma_g t)~,
\end{equation}
where the $Y_\ell^m$ are the usual spherical harmonic functions
\cite{as72}. For a given set of angular and azimuthal quantum numbers
$(\ell,m)$, the linearized adiabatic non-radial oscillation equations
reduce to a one-dimensional linear eigenvalue problem for $\sigma_g$ and
$\xi_r$, described by the following set of equations:
\begin{equation}
\frac{1}{r^2} \frac{d}{dr}(r^2 \xi_r)-\frac{g}{c_s^2} \xi_r +
\left(1-\frac{L_{\ell}^2}{\sigma_g^2}\right) \frac{P'}{\rho c_s^2}=
\frac{\ell (\ell+1)}{\sigma_g^2 r^2} \Phi' ,
\label{eq-osc1}
\end{equation}
\begin{equation}
\frac{1}{\rho} \frac{d P'}{dr}+\frac{g}{\rho c_s^2} P' +
(N^2-\sigma_g^2)\xi_r = -\frac{d\Phi'}{dr} ,
\label{eq-osc2}
\end{equation}
\begin{equation}
\frac{1}{r^2}\frac{d}{dr}\left(r^2\frac{d\Phi'}{dr}\right)-
\frac{\ell(\ell+1)}{r^2}\Phi' =
4 \pi G \rho \left(\frac{P'}{\rho c_s^2}+\frac{N^2}{g}\xi_r \right) ,
\label{eq-osc3}
\end{equation}
where $\Phi'$ is the perturbation of the gravitational potential, $c_s$ is
the sound speed, $L_{\ell}^2 \equiv \ell(\ell+1) c_s^2/r^2$ is the Lamb or
acoustic frequency, and $\xi_r$ is the (small) radial displacement
associated with a given mode of frequency $\sigma_g$ (see \cite{unn89} for
a detailed derivation). The eigenmodes associated with a given set of
$(\ell,m)$ values possess radial harmonics which can be labeled with a
third quantum number ($k$) related to the number of nodes in the
corresponding radial eigenfunction, so that the frequencies of individual
eigenmodes are best labeled as $\sigma_{k\ell m}$.

Inverting a continuous function, in our case $N^2(r)$, from a discrete set
of data (the pulsation periods) is well known to be a mathematically
ill-posed problem \cite{cb86,par94}. However, the situation is not as
critical as one might imagine because strong physical constraints can be
placed on the variations with depth of the Brunt-V\"ais\"al\"a frequency.
In white dwarf interiors, the $N^2(r)$ profile is determined by the
structural stratification (e.g., variations of density and pressure with
depth), which in turn depends on the star's evolutionary history as well
as a number of physical parameters such as stellar mass, core chemical
composition, surface temperature, and the mass of its surface helium
layer, to name but a few. The ill-posed inverse problem for $N^2$ can be
then recast in the form of an optimization problem that consists in
finding the numerical values for the set of these parameters that yields
the optimal fit between the oscillation periods of the corresponding white
dwarf structural model, as computed via the forward procedure outlined
above, and the observed periods. From the point of view of numerical
optimization, this is now a well-posed problem.

With detailed observations and a theoretical model in hand, the next step
is to select a suitable numerical optimization method. Models of all but
the simplest physical systems are typically non-linear, so finding the
optimal match to the observations requires an initial guess for each
parameter. Some iterative method is generally used to improve on this
first guess until successive iterations do not produce significantly
different answers. There are at least two potential problems with this
standard approach to model-fitting. The first guess is often derived from
the past experience of the person who is fitting the model. This {\it
subjective} method is even worse when combined with a {\it local} approach
to iterative improvement.  Many optimization schemes, such as differential
corrections \cite{pl72} or the simplex method \cite{kl87}, yield final
results that depend to some extent on the choice of initial model
parameters. This does not have serious consequences if the parameter-space
contains a single, well-defined minimum. But if there are many local
minima, then it can be more difficult for a traditional approach to find
the globally optimal solution (e.g., see Fig.~1 of \cite{cha95}).

The multi-modal nature of the optimization problem is not the only
modeling pitfall to be reckoned with. A good fit between model periods and
data certainly suggests that the model adequately reflects the actual
physical structure of the stars themselves. However, the possibility can
never be ruled out that other physical characteristics of the white dwarf
models, considered known and held fixed in the present modeling work,
could also be varied to yield comparably good fits to the observed
frequencies. As with any inverse problem, asteroseismic inferences are
plagued by the potential for non-uniqueness of the solutions. With this
caveat firmly in mind, we proceed.

\section{\bf Genetic Algorithms \label{GASEC}}

An optimization scheme based on a genetic algorithm (GA) can avoid the
problems inherent in many traditional approaches. The range of possible
values for each parameter is restricted only by observations and by the
constitutive physics of the model. Although the parameter-space defined in
this way is often quite large, a GA provides a relatively efficient means
of searching globally for the optimal model. Although it is more difficult
for GAs to find {\it precise} values for the optimal set of parameters
efficiently, they are well suited to search for the {\it region} of
parameter-space that contains the global minimum. In this sense, the GA is
an objective means of obtaining a good first guess for a more traditional
local hill-climbing method, which can narrow in on the precise values and
uncertainties of the optimal solution.

Genetic algorithms \cite{gol89,dav91,hol92,mit96}, arguably still the most
popular class of evolutionary algorithms \cite{mic96,bae96}, were inspired
by Charles Darwin's notion of biological evolution through natural
selection \cite{dar59}. The basic idea is to solve an optimization problem
by {\it evolving} the global solution, starting with an initial set of
purely random guesses. The evolution takes place within the framework of
the model, with the individual parameters serving as the genetic building
blocks. Selection pressure is imposed by some goodness-of-fit measure
between model and observations. Several books have been written to
describe how these ideas can be applied in a computational setting
\cite{gol89,dav91}, but we provide a basic overview below.

To begin, the GA samples the parameter-space at a fixed number of points
defined by a uniform selection of randomly chosen values for each
parameter. The GA evaluates the model for each set of parameters, and the
predictions are compared to observations. Each point in the ``population''
of trials is subsequently assigned a {\it fitness} based on the relative
quality of the match. A new generation of sample points is then created by
selecting from the current population of points according to their
computed fitness, and then modifying their defining parameter values with
two genetic operators in order to explore new regions of parameter-space.

Rather than modifying the parameter values directly, the genetic operators
are applied to encoded representations of the parameter sets. The simplest
way to encode them is to convert the numerical values of the parameters
into a string of digits. The string is analogous to a chromosome, and each
digit is like a gene. For example, a point defined by two parameters with
numerical values $a_1=0.123$ and $b_1=0.456$ could be encoded into the
string {\tt 123456}.

The two basic genetic operators are {\it crossover}, which emulates sexual
reproduction, and {\it mutation} which emulates somatic defects. The
crossover procedure pairs up the strings, chooses a random position for
each pair, and swaps the two strings from that position to the end. For
example, suppose that the encoded string above is paired with another
point having $a_2=0.567$ and $b_2=0.890$, which encodes to the string {\tt
567890}. If the second position between numbers on the string is chosen,
the strings become: 
\begin{eqnarray} 
{\tt 12}\fbox{\tt 3456} \rightarrow {\tt 12}\fbox{\tt 7890} \\ 
{\tt 56}\fbox{\tt 7890} \rightarrow {\tt 56}\fbox{\tt 3456} 
\end{eqnarray} 
To help keep favorable combinations of parameters from being eliminated or
corrupted too hastily, this operation is not applied to all of the pairs.
Instead, it is assigned a fixed occurrence probability ($p_c$) per
selected pair.

The mutation operator spontaneously replaces a digit in the string with a
new randomly chosen value. In our above example, if the mutation operator
is applied to the fourth digit of the second string, the result might be
\begin{equation}
{\tt 563}\fbox{\tt 4}{\tt 56} \rightarrow {\tt 563}\fbox{\tt 2}{\tt 56}
\end{equation}
Such digit replacement occurs with a small probability ($p_m$), often
dubbed the {\it mutation rate}.

After both operators have been applied, the strings are decoded back into
sets of numerical values for the parameters. In this example, the new
first string {\tt 127890} becomes $a'_1=0.127$ and $b'_1=0.890$, and the
new second string {\tt 563256} becomes $a'_2=0.563$ and $b'_2=0.256$. Note
that mutation in this case has caused a significant ``jump'' in
parameter-space, from the value $b'_2=0.456$ that would have been
generated by the crossover operation only. The new genetically-shuffled
set of points replaces the original set, and the process is repeated until
some termination criterion is met.

What is the justification for this rather contorted way to produce two new
trial points from two existing ones? One could have instead simply formed
the arithmetic averages of the pairs $a_1,a_2$, and $b_1,b_2$. However,
under the ``pressure'' of fitness-based selection, crossover acting on
successive generations of strings modifies the frequency of a given
substring in the population at a rate proportional to the difference
between the mean fitness of the subset of strings incorporating that
substring, and the mean fitness of all strings making up the current
population. This mouthful is given quantitative expression in the
so-called {\it schema theorem} \cite{gol89,hol92,mit96}, which continues
to form the basis of most theoretical analysis of GAs. The GA can be
thought of as a classifier system that continuously sorts out and combines
the most advantageous substrings that happen to be present across the
whole population at a given time.\footnote{In the GA literature this
property is known as ``intrinsic parallelism'' \cite{hol92}, which has
nothing to do with the practical issue of implementing a GA application on
a parallel hardware architecture.} In this context the role of mutation is
to inject ``novelty'' continuously, by producing new digit values at
specific string positions, which might not otherwise have been present in
the population or may have been selected against during earlier
evolutionary phases.

It should be clear already from this brief introductory discussion that
the operation of a GA involves a number of random processes, so that the
resulting search algorithm is stochastic in nature. Consequently, there is
always a finite probability that the GA will not find the globally optimal
solution in a given run. This probability decreases gradually, of course,
as the evolution is pushed through more and more generations. Alternately,
one can run the GA for fewer generations, but do so several times with
different random initialization. This form of higher-level Monte Carlo
simulations makes it possible to establish the validity of the optimal set
of model parameters with an acceptable degree of confidence.

\section{\bf The PIKAIA subroutine}

{\tt PIKAIA} is a self-contained, genetic-algorithm-based optimization
subroutine developed at the High Altitude Observatory, and available in
the public domain
(http://www.hao.ucar.edu/public/research/si/pikaia/pikaia.html). {\tt
PIKAIA} maximizes a user-specified FORTRAN function through a call in the
body of the main program. Unlike many GA packages available commercially
or in the public domain, {\tt PIKAIA} uses decimal (rather than binary)
encoding. This choice was motivated by portability issues---binary
operations are usually carried out through platform-dependent functions in
FORTRAN, which makes it more difficult to port the code between PC and
workstation platforms. While originally designed primarily as a learning
tool, {\tt PIKAIA}'s portability, ease of use, and robustness have made it
by all appearances the software of choice for a wide variety of modeling
problems requiring global optimization capabilities (see
\cite{cha98,hpe99,hh00,tk01} for sample applications; and the {\tt PIKAIA}
web page for a compilation of past and present users and their research
applications).

\begin{figure}[p]
\epsfxsize 5.0in
\epsffile{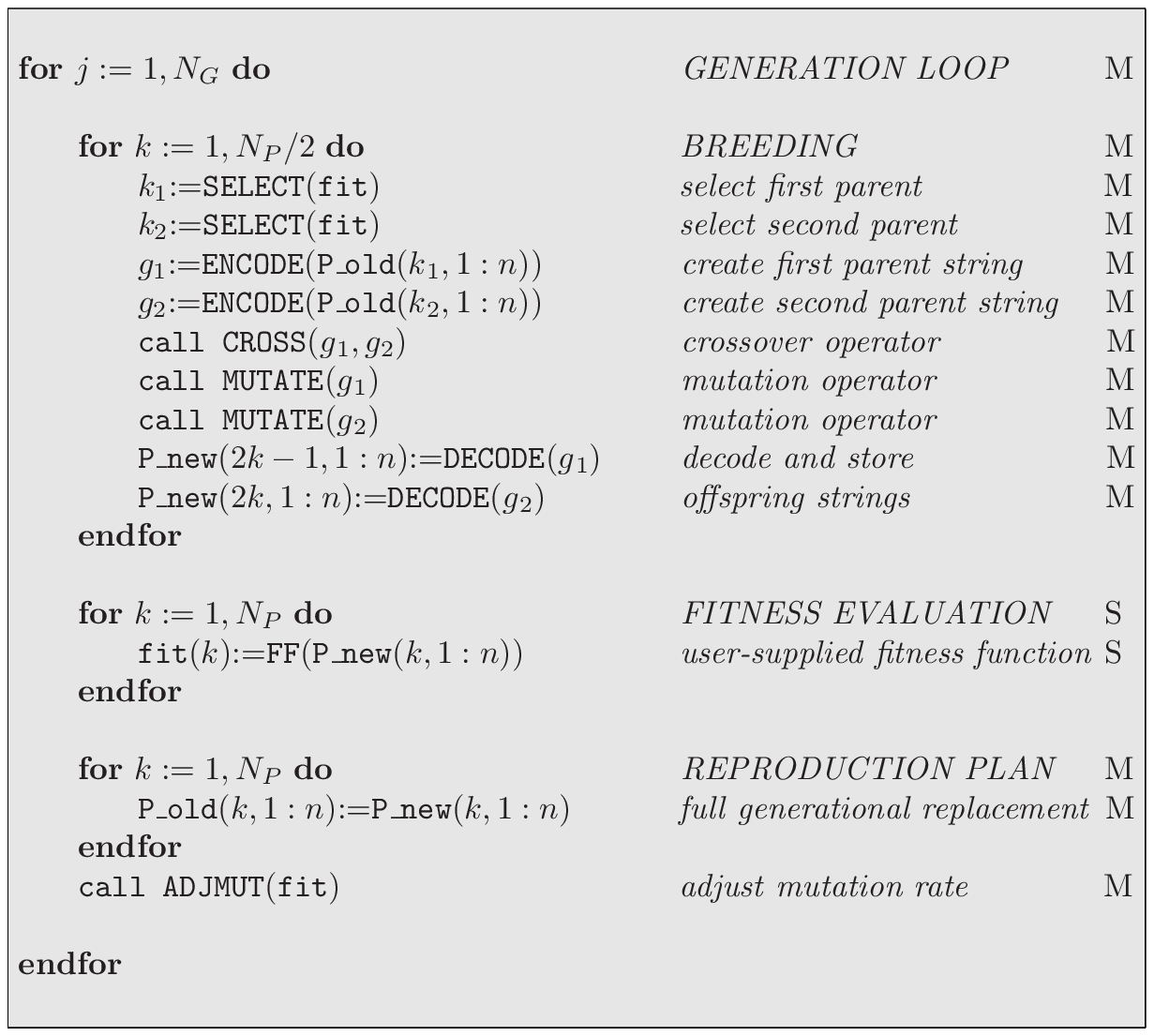}
\caption{\footnotesize\ Minimal pseudo-code for the {\tt PIKAIA} genetic 
algorithm optimization subroutine, operating in Full-Generational-Replacement 
mode. The number of model parameters being optimized is $n$, $N_P$ is the
(fixed) population size, and $N_G$ is the number of (time-like)
generations over which the evolution is carried out. The functions {\tt
ENCODE} and {\tt DECODE} convert a $n$-dimensional floating-point array to
a string, and vice-versa. The function {\tt SELECT} picks a single
individual (flagged by an integer, here $k_1$ or $k_2$), with a
probability proportional to its fitness-based rank in the current
population. Note that the strings $g_1$, $g_2$ are modified upon exit from
{\tt CROSS} and {\tt MUTATE}, and since each breeding event produces two
offspring, the first inner loop only needs to repeat $N_P/2$ times to
produce a new, full-size population. Operations and function calls labeled
``M'' are carried out serially by the Master program, and those labeled
``S'' are executed in parallel by the Slave processes (see \S
\ref{APPSEC} for more details).\label{fig1}}
\end{figure}

\subsection{GA structure: Select-Breed-Evaluate-Replace}

Fig.~\ref{fig1} shows, in pseudo-code form, the algorithmic structure of
{\tt PIKAIA}. The inner workings of the various functions and subroutines
appearing therein have been described at length elsewhere
\cite{cha95,chk95}, and so are only outlined in what follows. We do
describe in some detail in \S \ref{adjmutsec} and \S \ref{creepsec} below
additional strategies and operators not originally included in the
public-release version {\tt PIKAIA 1.0}.

The task at hand is the {\it maximization} of the user-supplied function
{\tt FF}, which accepts as input an $n$-dimensional floating-point array
{\tt x}($1:n$) containing a set of parameter values defining one instance
of the model being optimized, and returning a measure of goodness-of-fit
(based, e.g., on a $\chi^2$ measure if the model output is being compared
to data).  The code evolves a population of $N_P$ trial points in the
$n$-dimensional search space, stored in the array {\tt
P\_old}($1:N_P,1:n$), through a preset number of generations $N_G$. The
population is (usually) initialized with random deviates uniformly
distributed in user-specified intervals defining the range of
parameter-space to be explored, so that the evolutionary search remains
bounded but otherwise entirely unbiased by the choice of initial
conditions.

At each time-like generation (outer loop), pairs of ``parents'' are
extracted from the current population, with selection probability
increasing with the individual's fitness, using a rank-based version of
the classical Roulette Wheel Algorithm \cite{dav91}. The two corresponding
$n$-dimensional floating-point arrays are then encoded into two strings
($g_1,g_2$), bred using crossover and mutation operators, and decoded back
into two $n$-dimensional floating-point arrays that define the two
``offspring'' points. These are stored in the temporary array {\tt
P\_new}, which concludes the {\it breeding} step. The fitness of the new
population members is then computed via the user-supplied fitness function
{\tt FF}, and stored in the $N_P$-dimensional array {\tt fit}. Finally, a
{\it reproduction plan} is needed to insert some or all of the newly
generated and evaluated trial points into the breeding population. The
simplest strategy, dubbed ``Full-Generational-Replacement'', consists in
breeding a number of new trial points equal to that within the original
population (first inner loop, repeating only $N_P/2$ times since each
breeding event produces two offspring), and then replacing the old
population with the new (third inner loop). This concludes one
generational iteration, and the above steps are repeated anew.

\subsection{Dynamical adjustment of the mutation rate \label{adjmutsec}}

Of the various internal parameters governing the operation of the genetic
algorithm itself, the {\it mutation rate} is one that often sensitively
affects performance \cite{gol89,bae96}. This rate is more aptly defined as
the probability ($0\leq p_m\leq 1$) that a single digit in the encoded
strings will be subjected to replacement by another random digit, as
already described briefly in \S \ref{GASEC}. As with biological systems,
mutation is very much a mixed blessing. It represents the only way to
inject variability into the evolving population \cite{hol92}, which is
extremely useful---some would say essential---for efficient exploration of
parameter-space and displacement of the population away from secondary
extrema. But too much mutation can also destroy the existing good
solutions. Some sort of optimal tradeoff between these incompatible
tendencies must be achieved by a proper choice of $p_m$.

Various ``recipes'' for setting $p_m$ have been put forth, starting with
simple prescriptions such as setting $p_m=(N_PL)^{-1}$, where $L$ is the
string length \cite{DeJ75}, detailed empirical modeling \cite{mh01}, all
the way to meta-simulations where a higher-level GA evolves the
combination of controlling parameters that yields the best performance of
the underlying GA for the problem under consideration \cite{gref86}.
However, near-optimal performance is rarely sustained across wide ranges
of problems. Moreover, because of the very large number of model
evaluations involved, some of the more elaborate approaches rapidly become
impractical if the modeling problem at hand is very computation intensive,
as is the case with the problem considered here.

One way around this quandary is to allow $p_m$ to vary dynamically in the
course of an evolutionary run, according to the degree of clustering of
the current population as a whole. The public-release version of {\tt
PIKAIA} does so by monitoring the normalized difference ($\Delta$) between
the fitness values of the best and median individuals in the population
(ranking being based on fitness):
\begin{equation}
\Delta={\max({\tt fit})-{\rm med}({\tt fit})
\over\max({\tt fit})+{\rm med}({\tt fit})},
\qquad [{\rm fitness~based}].
\label{Eq-adj1}
\end{equation}
A strongly clustered (scattered) population has $\Delta\to 0$ ($\Delta\to
1$). At the end of each generational iteration, $\Delta$ is computed, and
if found to fall below (exceed) a preset lower (upper) bound $\Delta_1$
($\Delta_2$), the mutation rate is multiplicatively incremented
(decremented) by a factor $\delta$:
\begin{equation}
p_m\to \cases{p_m\times \delta,  & $\Delta\leq \Delta_1$,\cr
                p_m/\delta, & $\Delta\geq \Delta_2$.\cr}
\label{Eq-adj2}
\end{equation}
Experience reveals that the GA's performance is not sensitively dependent
on the adopted values for $\Delta_1, \Delta_2$ and $\delta$, within
reasonable bounds. Numerical values $\Delta_1=0.05$, $\Delta_2=0.25$,
$\delta=1.5$ have proved to be robust over a variety of test problems, and
are hardwired in {\tt PIKAIA}'s mutation rate adjustment subroutine ({\tt
ADJMUT} in Fig.~\ref{fig1}).

\begin{figure}[t]
\hskip 0.375in
\epsfxsize 4.0in
\epsffile{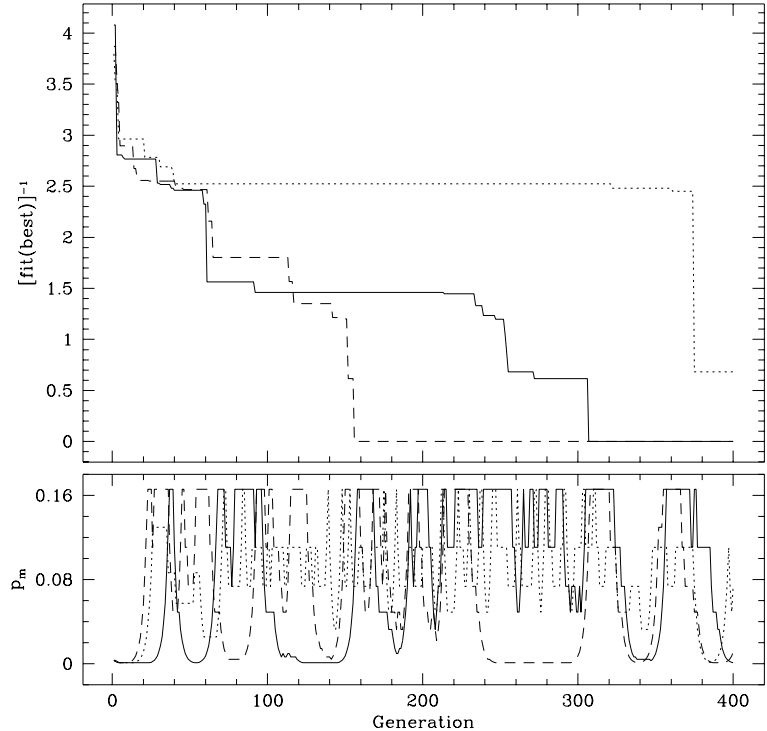}
\caption{\footnotesize\ Sample convergence curves for GA runs using different 
methods for dynamical adjustment of the mutation rate. The top panel shows 
the variance of the best trial in the population as a function of generation
using distance-based adjustment with (solid) and without (dashed) creep
mutation included, and using fitness-based adjustment with creep mutation
(dotted). The bottom panel shows the corresponding mutation rate as a
function of generation for the three methods. Crossover is primarily
responsible for the initial rapid improvement in all three curves. Both
distance-based curves converge to the global solution, though more slowly
when creep mutation is included. The fitness-based method is much less
successful escaping local minima, but eventually converges to the region
of the global solution.\label{fig2}}
\end{figure}

Clearly, Eq.~(\ref{Eq-adj1}) is not the only possible measure of
population clustering. Another possibility is to use a measure of metric
distance between the best and median individual in the population:
\begin{equation}
\Delta^2
={1\over n}\sum_{k=1}^n (x_k^{\rm max}-x_k^{\rm med})^2
\qquad [{\rm distance~based}],
\label{Eq-adj3}
\end{equation}
where $x_k^{\rm max}$ ($x_k^{\rm med}$) represents the $k^{\rm th}$
element of the $n$-dimensional floating-point array containing the
parameter values defining the current best (median) individual in the
population. Which of Eqs.~(\ref{Eq-adj1}), (\ref{Eq-adj3}) will yield the
best optimization performance cannot be foretold, as the answer will
depend on the shape of the fitness isosurface in parameter-space (and if
these are known {\it a priori} in detail, then the optimization problem is
already solved!). On synthetic white dwarf data, distance-based adjustment
was found to increase the success rate (i.e., probability of locating the
true global optimum) of the search process by $\sim$50\% over
fitness-based adjustment, all other GA control parameters being the same
(see Fig.~\ref{fig2}).

Dynamical adjustment of $p_m$ can lead to relatively large mutation rates
in some evolutionary phases, with the potential danger of destroying good
solutions.  {\tt PIKAIA} avoids this by making use of a small but very
useful ``cheat'' known as {\it elitism} \cite{DeJ75}, which saves the
fittest member of the breeding population ({\tt P\_old}) and recopies it
to the new population ({\tt P\_new}) as the final step of the reproduction
plan.

{\tt PIKAIA}'s dynamical adjustment of the mutation rate represents a
particularly simple form of {\it self-adaptation} (see \cite{bae01}, and
articles therein). {\tt PIKAIA} can function anywhere along a spectrum
extending between two very qualitatively distinct search modes; as long as
$p_m\ll 1$, {\tt PIKAIA} operates as a classical genetic algorithm, with
the crossover operator primarily responsible for the exploration of
parameter-space. On the other hand, as $p_m\to\Delta_2$ {\tt PIKAIA}
behaves more and more like a stochastic iterated hill-climber. This
peculiar algorithmic combination has proven to be both robust and
efficient.

\subsection{Creep mutation \label{creepsec}}

The one-point mutation operator included in the public-release version of
{\tt PIKAIA} suffers from a well-known shortcoming, which may prove
troublesome under certain problem-dependent circumstances. Consider the
following portion of a string encoding a parameter value $a=0.1961$:
\begin{equation}
{\tt ...1961...}
\end{equation}
Assume now that the optimal solution has $a^\ast=0.2050$, and that the
search space is smooth enough that, at least early in the evolution, an
individual with $a=0.1961$ has above-average fitness. The occurrence
frequency of the above substring in the population will increase, and
sequential action of one-point mutation is likely to lead to a gradual
trend toward the substring
\begin{equation}
{\tt ...1999...}
\end{equation}
But now we have a problem. Going from this substring to the optimal {\tt
2050} requires some simultaneous and well-coordinated digit substitutions
on the part of one-point mutation, which, statistically, are highly
unlikely. The string has become stuck at a so-called ``Hamming wall''.

Encoding schemes can be designed such that successive single digit changes
in the string translate into smooth variations of the decoded parameter
values. Binary Gray coding \cite{mic96,ptvf92} is a well-known example.  
An alternate, often more practical solution is to make use of a new
mutation operator known as {\it creep mutation} \cite{dav91}. In the
context of decimal encoding, this would operate as follows. Rather than
randomly replacing a digit targeted for mutation, add or subtract $1$
(with equal probabilities) to the existing digit, and ``carry over the
one'' when appropriate. For example,
\begin{equation}
{\tt ...19}\fbox{\tt 9}{\tt 9...}\rightarrow 
{\tt ...1}\fbox{\tt 90}{\tt 9...}\rightarrow 
{\tt ...}\fbox{\tt 100}{\tt 9...}\rightarrow 
{\tt ...2009...}.
\end{equation}
Clearly creep mutation can cross Hamming walls, but it lacks the ability
to generate occasional large ``jumps'' in parameter-space the way
one-point mutation can if it operates on a string element that decodes
into a leading digit in the corresponding floating-point parameter. Since
this latter property is advantageous for exploration of parameter-space,
whenever carrying out mutation ({\tt call MUTATE} in Fig.~\ref{fig1}), it
is preferable to pick either one-point or creep mutation with equal
probabilities.

The use of creep mutation for our white dwarf modeling problem gave mixed
results; it led to a higher probability of finding the exact set of
optimal parameters, but at the expense of slightly slower convergence to
the region of the global solution (see Fig.~\ref{fig2}). This probably
reflects the fact that there were no Hamming walls in the vicinity of the
optimal solution. In other {\tt PIKAIA} applications, however, creep
mutation has been found to lead to significant improvements. As with so
many other aspects of GA-based numerical optimization, the benefit of
creep mutation is highly problem-dependent.

\section{\bf Parallel Implementation \label{APPSEC}}

In 1998 we began a project to adapt some well-established white dwarf
evolution and pulsation codes to interface with {\tt PIKAIA}. On the
fastest processors available at the time, a single model would run in
about 45 wallclock seconds. Knowing that the optimization would require
$\sim$10$^{\rm 5-6}$ models, it was clear that a serial version of {\tt
PIKAIA} would require many months to finish on a single processor. So we
decided to incorporate the message passing routines of the Parallel
Virtual Machine (PVM) software \cite{gei94} into {\tt PIKAIA}.

\subsection{General parallelization considerations}

The PVM software allows a collection of networked computers to cooperate
on a problem as if they were a single multi-processor parallel machine.
All of the software and documentation is free. We had no trouble
installing it on our Linux cluster \cite{mn00} and the sample programs
that come with the distribution made it easy to learn and use. The
trickiest part of the procedure was deciding how to split up the workload
among the various computers.

\begin{figure}[t]
\epsfxsize 5.0in
\epsffile{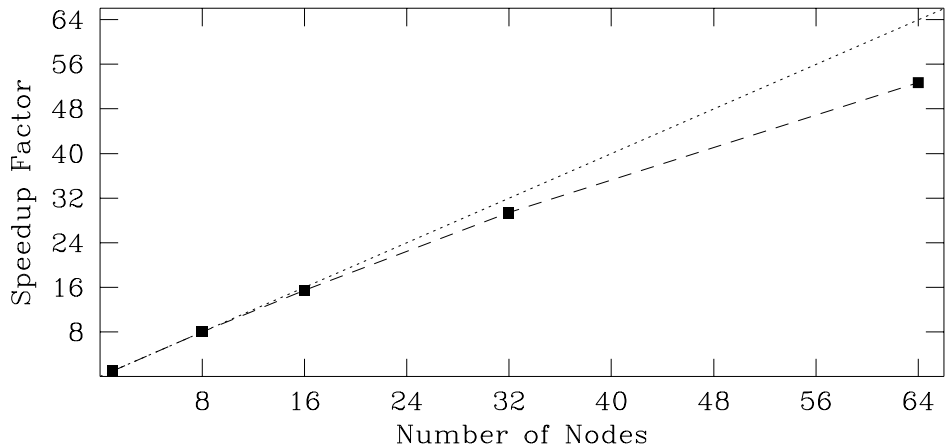}
\caption{\footnotesize\ The scalability of the parallel version of 
{\tt PIKAIA} on a 64-node Linux cluster using the white dwarf fitness 
function. The cluster contains two types of CPUs (PIII and K6-2) running 
at slightly different speeds (300 and 366 MHz), but the points in this 
plot were produced using equal numbers of each, scaled relative to the 
average performance of the two types.\label{fig3}}
\end{figure}

The GA-based fitting procedure for the white dwarf code quite naturally
divided into two basic functions: evolving and pulsating white dwarf
models, and applying the genetic operators to each generation once the
fitnesses had been calculated. When we profiled the distribution of
execution time for each part of the code, this division became even more
obvious. Here, as with the vast majority of real-life applications,
fitness evaluation (second inner loop on Fig.~\ref{fig1}) is by far the
most computationally demanding step. For our model-fitting application,
93\% of CPU time is spent carrying out fitness evaluation, 4\% carrying
out breeding and GA internal operations (such as mutation rate
adjustment), and 3\% for system and I/O. It thus seemed reasonable to
create a {\it slave} program to perform the model calculations, while a
{\it master} program took care of the GA-related tasks.

In addition to decomposing the function of the code, a further division
based on the data was also possible. Fitness evaluation across the
population is inherently a parallel process, since each model can be
evaluated independently of the others. Moreover, it requires minimal
transfer of information, since all that the user-supplied function {\tt
FF} requires is the $n$-dimensional floating-point array of parameters
defining one single instance of the model, and all it needs to return is
the floating-point value corresponding to the model's fitness.  It is then
natural to send one model to each available processor, so the number of
machines available would control the number of models that could be
calculated in parallel. Maximal use of each processor is then assured by
choosing a population size $N_P$ that is an integer multiple of the number
of available processors.

In practice, this recipe for dividing the workload between the available
processors proved to be very scalable. Since very little data is exchanged
between the master and slave tasks, our 64-node cluster provided a speedup
factor of about 53 over the performance on a single processor (see
Fig.~\ref{fig3}).

\subsection{Master Program}

\begin{figure}[p]
\hskip 0.5in
\epsfxsize 4.5in
\epsffile{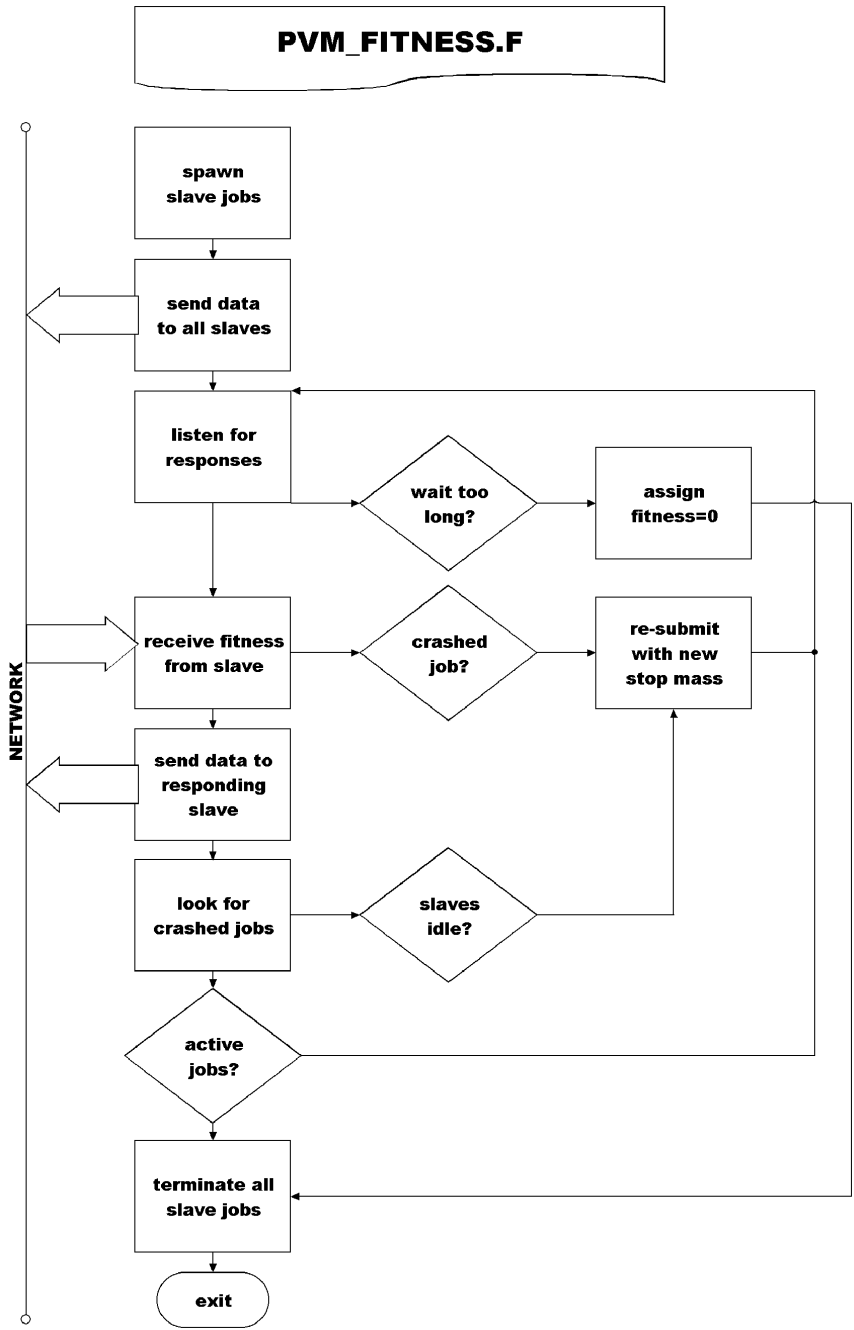}
\caption{\footnotesize\ Flow chart for the parallel fitness evaluation 
subroutine, which runs on the master computer.\label{fig4}}
\end{figure}

Starting with the slightly improved unreleased version of {\tt PIKAIA},
including creep mutation and distance-based mutation rate adjustment, we
used the message passing routines from PVM to create a parallel fitness
evaluation subroutine. The original code evaluated the fitnesses of the
population one at a time in a DO loop (equivalent to the second inner loop
in Fig.~\ref{fig1}). We replaced this procedure with a single call to a
new subroutine that evaluates the fitnesses in parallel on all available
processors. The parallel version of {\tt PIKAIA} constitutes the master
program, which runs on the central computer of our Linux cluster. A flow
chart for the parallel fitness evaluation subroutine ({\tt
PVM\_FITNESS.F}) is shown in Fig.~\ref{fig4}.

After starting the slave program on every available processor, {\tt
PVM\_FITNESS.F} sends an array containing scaled values of the parameters
to each slave job over the network. In the first generation of the GA,
these values are completely random; in subsequent generations, they are
the result of the selection, crossover, and mutation of the previous
generation, performed by the non-parallel portions of {\tt PIKAIA}.

Next, the subroutine listens for responses from the network and sends a
new set of parameters to each slave job as it finishes the previous
calculation. When all sets of parameters have been sent out, the
subroutine begins looking for jobs that may have crashed and re-submits
them to slaves that have finished and would otherwise sit idle. If a few
jobs do not return a fitness after an elapsed time much longer than the
average runtime required to compute a model, the subroutine assigns them a
fitness of zero. In a typical run, this was necessary for less than 1 in
10,000 model evaluations. When every set of parameters in the generation
have been assigned a fitness value, the subroutine returns to the main
program to perform the genetic operations---resulting in a new generation
of models to calculate. The process continues for a fixed number of
generations, chosen to maximize the success rate of the search.

In all simulation runs reported below, we kept the crossover probability
fixed at $p_c=0.85$, used an initial mutation probability $p_m=0.005$, and
retained a level of selection pressure corresponding to {\tt PIKAIA}'s
default settings. We determined the optimal number of generations by
applying the method to synthetic data and looking at the fraction of runs
that converged to the input model as a function of run length, up to 500
generations. To minimize the probability of missing the global solution
for observed data, we fixed the number of generations to be slightly
larger than the test run that converged the slowest. Within the range of
control parameters we explored, our GA application exhibited a clear
tradeoff effect between population size $N_P$ and number of generational
iterations $N_G$. As long as the product $N_PN_G$ remained near $3\times
10^4$, the success probability of a single GA run was approximately
constant at $\sim$70\%.

\subsection{Slave Program}

The original white dwarf code came in three pieces: (1) the evolution
code, which starts with a static polytropic approximation of a pre-white
dwarf and allows it to cool quasi-statically until it reaches the desired
temperature, (2) the prep code, which reformats the output of the
evolution code, and (3) the pulsation code, which uses the output of the
prep code to solve the adiabatic non-radial oscillation equations,
yielding the mode periods to be compared with the observed periods.

\begin{figure}[t]
\hskip 1.0in
\epsfxsize 3.5in
\epsffile{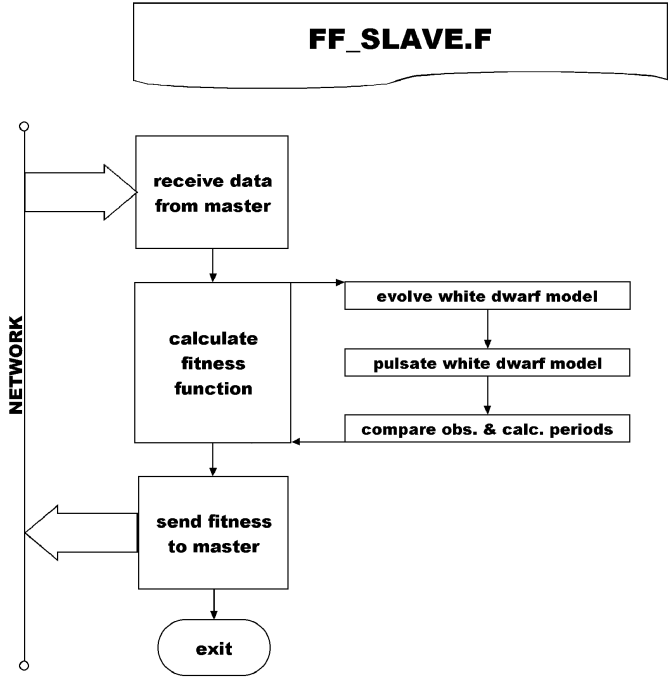}
\caption{\footnotesize\ Flow chart for the slave program of the parallel 
code, which runs on each node of the Linux cluster.\label{fig5}}
\end{figure}

To get the white dwarf code running in an automated way, we merged the
three components of the original code into a single program, and added a
front end that communicated with the master program through PVM routines.
This code ({\tt FF\_SLAVE.F}) combined with the fitness function
constitutes the slave program, and is run on each node of the Linux
cluster. A flow chart for {\tt FF\_SLAVE.F} is shown in Fig.~\ref{fig5}.

The operation of the slave program is relatively simple. Once it is
started by the master program, it receives a set of parameters from the
network. It then calls the fitness function (the white dwarf code in our
case) with these parameters as arguments. Our fitness function rescales
the dimensionless parameters into physical units, and cools a polytropic
white dwarf model with the proper mass and structure down to the specified
temperature. A typical model has several hundred spherical mass shells,
and requires at most a few dozen time steps to cool down to the relevant
temperatures.

The slave program then calculates the adiabatic non-radial pulsation
periods within a specified range, given the spherical degree of the modes
(only $\ell\le2$ have been observed in white dwarfs). This involves
solving Eqs.~(\ref{eq-osc1})--(\ref{eq-osc3}), which is carried out
numerically using an iterative scheme based on the Runge-Kutta-Fehlberg
shooting method \cite{hk94}. The first guess for the eigenvalue can be
obtained from the following useful approximation of the $g$-mode pulsation
frequencies:
\begin{equation}
  \sigma_{k\ell m} \approx 
    \left< {N^2 \ell(\ell+1) \over k^2 r^2} \right> ^{1/2}
  + \left( 1 - {C_k \over \ell(\ell+1)} \right) m\Omega
\end{equation}
where the second term is due to the slow rotation frequency $\Omega$
(which breaks the spherical symmetry), and the constant $C_k$ is of order
unity \cite{win98}. Typically, only a few iterations are needed to achieve
convergence on a radial mesh with several thousand effective zones, where
the model quantities are interpolated by means of a cubic spline between
the equilibrium model shells, equally spaced in mass.

Finally, each of the observed periods ($P_{\rm obs}$) are compared to the
nearest model periods ($P_{\rm mod}$), and the variance ($\sigma$) is
calculated,
\begin{equation}
\sigma=\left({1\over N}\sum_{j=1}^N (P_{\rm obs}-P_{\rm mod})^2\right)^{1/2}~,
\end{equation}
with $N=11$ for the data used here. The fitness is then defined as the
inverse of this root-mean-square period residual\footnote{The exact
definition of fitness in terms of $\sigma$ is not critical here, since
{\tt PIKAIA} establishes selection probability in terms of fitness-based
{\it ranks}. In other words, defining fitness as $1/\sigma^2$ instead of
$1/\sigma$ would lead to the same selection probabilities in a given
population of trial solutions.}, and is sent to the master program over
the network. The node is then ready to run the slave program again and
receive a new set of parameters from the master program.

\section{\bf Results}

The ultimate goal of our project was to derive a measurement of the
astrophysically important $^{12}{\rm C}(\alpha,\gamma)^{16}{\rm O}$
nuclear reaction rate. When a white dwarf star is being formed in the core
of a red giant during helium burning, the $3\alpha$ and $^{12}{\rm
C}(\alpha,\gamma)^{16}{\rm O}$ reactions compete for the available helium
nuclei. The relative rates of the two reactions determines the final yield
of oxygen deep in the core. The $3\alpha$ rate is well established, but
the same is not true of the $^{12}{\rm C}(\alpha,\gamma)^{16}{\rm O}$
reaction. The extrapolation of its rate to stellar energies from
high-energy laboratory measurements is complicated by interference between
various contributions to the total cross-section, leading to a relatively
large uncertainty \cite{kun01}. This translates into similarly large
uncertainties in our understanding of every astrophysical process that
depends on this reaction, from supernovae explosions to galactic chemical
evolution. A seismological measurement of the core oxygen mass fraction
$X_{\rm O}$ in a pulsating white dwarf star can provide an independent way
to determine the $^{12}{\rm C}(\alpha,\gamma)^{16}{\rm O}$ reaction at
stellar energies.

Knowledge of the central oxygen mass fraction has other important
astrophysical implications. Surveys of white dwarfs in our galactic
neighborhood have demonstrated a marked deficit at luminosities below
about $10^{-6}$ times the solar luminosity ($L_\odot$) \cite{ldm88}. The
favored interpretation is that even the oldest white dwarfs in the galaxy
have not yet had time to cool below the observed cutoff luminosity
\cite{wa87}. This opened the possibility to infer the age of the galactic
disk, and thus obtain a lower limit on the age of the Universe. White
Dwarf Cosmochronometry, as the subject has been called, evidently requires
detailed knowledge of the white dwarf thermal energy content and cooling
history \cite{fbb01}. The former turns out to depend significantly on the
core chemical composition, primarily the mass ratio of carbon-to-oxygen
($X_{\rm C}/X_{\rm O}$), these being the two constituents that
theoretically account for the near totality of the core mass.

Our first application of the parallel GA allowed only 3 parameters to be
varied in the models: the surface temperature, the stellar mass, and the
thickness of a surface helium layer. To the extent possible, we defined
the boundaries of the search using only observational constraints and
limits imposed by the underlying physics. It was the broadest survey of
white dwarf pulsation models ever conducted, covering more than 100 times
the search volume of previous studies. After demonstrating that the method
was successful on synthetic data, we applied it to the best-observed star
among the helium-atmosphere pulsators, GD~358, using data obtained by the
Whole Earth Telescope \cite{nat90}. The original analysis of these data
\cite{win94} identified a series of eleven ($\ell=1, m=0$) pulsation modes
of consecutive radial overtone ($k=8$--18) with periods between 400 and
900 seconds. The measured trigonometric parallax of GD~358 confirmed this
$\ell$ identification beyond doubt, since the luminosity of models with
higher $\ell$ modes could not be reconciled with this independent
constraint. The initial asteroseismic study of GD~358 from these data
concluded that the helium layer mass was near $10^{-6}~m/M_*$ \cite{bw94},
a result at odds with standard stellar evolution theory, which leads to an
expected value near $10^{-2}~m/M_*$ \cite{dm79}.

\begin{figure}[t]
\epsfxsize 5.0in
\epsffile{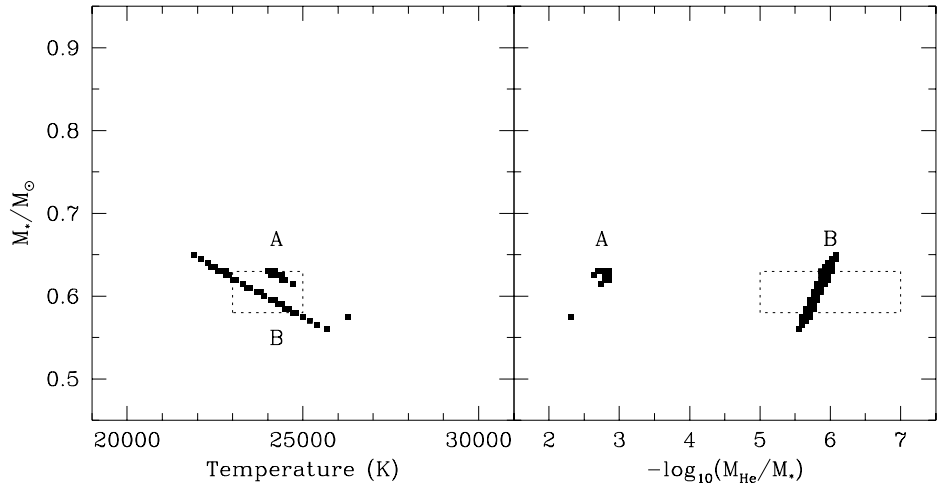}
\caption{\footnotesize\ Front and side views of the three-dimensional GA 
search space (stellar mass; surface temperature; surface helium layer mass) 
for a C/O 50:50 core with a ``steep'' internal chemical profile. Square 
points mark the locations of every model found by the GA with 
root-mean-square period residuals smaller than 3 seconds. The dotted line 
shows the range of parameters considered in \cite{bw94}. The best solution 
in family A has $\sigma=2.71$, while the best in family B has $\sigma=2.42$. 
Since the typical observational uncertainties in period determinations are 
about $\sim$0.05 seconds, the difference is statistically and physically 
significant. Note that the optimal solution for this core composition
belongs to family B, but when the internal composition and structure are
also optimized this is no longer the case (see \cite{mnw00,mwc01} for 
details).\label{fig6}}
\end{figure}

The global search performed by the parallel GA led to the discovery of two
families of reasonably good models for this object, with helium layer
masses near $10^{-6}$ and $10^{-2}~m/M_*$ \cite{mnw00}. Fig.~\ref{fig6}
shows front and side views of the complete 3-dimensional parameter-space
covered by our search, along with the two good families of models. The
dotted lines show the range of parameters covered by the earlier search.
When we confined the GA to search within the dotted region, it found a
solution consistent with that found by the earlier investigation. But the
family of models with thicker helium layers ultimately provided a better
fit to the observations ($\sigma = 1.5$ seconds \cite{mnw00}), resolving
the tension with stellar evolution theory. This discovery would not have
been possible if we had confined our search using more subjective
criteria.

Based on our initial success, we extended the method to include two
additional parameters to describe the internal composition and structure
of the white dwarf: the central oxygen mass fraction $X_{\rm O}$, and a
parameter related to the width of the composition transition layer between
the core and envelope of the white dwarf. After some initial difficulty,
we realized that two of our model parameters were correlated, so we
devised a system using the GA to iteratively optimize two sets of 4
parameters until both fits converged to the same point. With tests on
synthetic data, we demonstrated that the success probability was 70-80\%
for an individual run (with $N_G=400+250$, $N_P=128$). By selecting the
best solution from 10 independent runs, the probability of missing the
global solution became negligible ($<10^{-6}$). The entire optimization
procedure required 3 iterations between the subsets of 4 parameters. Each
iteration consisted of 10 runs with a total of 650 generations of 128
points. In the end, the method required $2.5\times10^6$ model evaluations,
which was 200 times more efficient than enumerative search of the grid for
each iteration at the same sampling density, or about 4,000 times more
efficient than enumerative search of the entire 5-dimensional space.

\begin{figure}[t]
\hskip 0.375in
\epsfxsize 4.0in
\epsffile{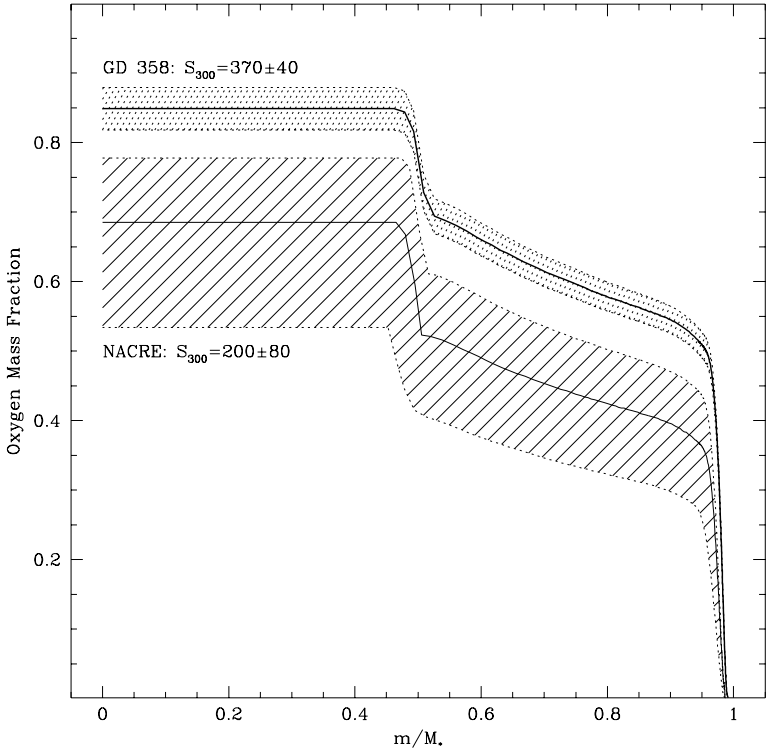}
\caption{\footnotesize\ The internal oxygen profiles for a $0.65~M_{\odot}$ 
white dwarf model using the NACRE \cite{ang99} nuclear reaction rates (solid 
line) between the upper and lower limits on the $^{12}{\rm C}(\alpha,\gamma)  
^{16}{\rm O}$ rate (hashed region). Also shown are the profiles resulting
from the rates that match the central oxygen mass fraction derived for
GD~358 (thick solid line)  within the $\pm 1\sigma$ limits (shaded
region) \cite{msw02}.\label{fig7}}
\end{figure}

The end result of our new fit included a measured value for the crucial
central oxygen mass fraction in GD~358: $X_{\rm O}=84\pm3$ percent
\cite{mwc01}. The age estimate of the galactic disk associated with white
dwarf cooling to $10^{-6}\,L_\odot$ has been shown to vary by as much as
3.6 Gyr as $X_{\rm O}$ varies from zero (pure carbon core) to unity (pure
oxygen core) \cite{fbb01}. If our measurement of $X_{\rm O}$ for GD358 is
characteristic of most white dwarfs, this would imply an age for the
galactic disk near the low end of the cosmochronologically allowed range
(8.5-11 Gyr; see \S 3.2 and Figs. 6 \& 7 in \cite{fbb01}).

Combined with detailed simulations of white dwarf internal chemical
profiles, our determination of $X_{\rm O}$ also makes it possible to infer
the $^{12}{\rm C}(\alpha,\gamma)^{16}{\rm O}$ reaction rate \cite{mwc01}.
Using an evolutionary model that produced the same final mass as GD~358,
the value of the $^{12}{\rm C}(\alpha,\gamma)^{16}{\rm O}$ rate was
adjusted until the central oxygen mass fraction matched our
asteroseismically inferred value (see Fig.~\ref{fig7}). The implied
reaction rate was $S_{300}=370\pm40$ keV b \cite{msw02}. Considering that
the root-mean-square period residuals of our optimal model are still
$\sim$1 second while the observational uncertainties are closer to
$\sim$0.05 second, there is clearly more work to be done. But this new
computational method will allow us to probe the fine details of white
dwarf interior structure that were formerly inaccessible, like the
detailed variation with depth of the interior composition.

Finally it is worth reiterating the non-uniqueness caveat mentioned in \S
\ref{INTROSEC}. While we are confident that the solutions obtained here
are optimal from the point of view of residual minimization, they are only
so {\it within our modeling framework}. We have reduced the possible
variations of the internal stratification of white dwarfs to five primary
parameters. In doing so we are sampling a small subset of the space of all
possible and (physically consistent) internal stratifications. In order to
draw firm conclusions from our results, we need to further assume that the
subset defining our search space is representative of the full space of
possible solutions. Encouraging evidence that this might well be the case
has already been obtained, by using the GA to evolve physically motivated,
local perturbations to the Brunt-V\"ais\"al\"a frequency profile that lead
to further statistically significant improvement in the period residuals
\cite{mwc01}. Nonetheless, the possibility still remains that even better
and physically distinct families of solutions lie somewhere in dimensions
of ``model space'' that we have not yet explored. This must be kept in
mind when making a final assessment of the physical conclusions described
above.

\section{\bf Discussion}

The application of genetic-algorithm-based optimization to white dwarf
pulsation models turned out to be very fruitful. We are now confident that
we can rely on this approach to perform global searches and to provide
objectively determined optimal models for the observed pulsation
frequencies of white dwarfs, along with fairly detailed maps of the
parameter-space as a natural byproduct. The method finally allowed us to
measure the central oxygen mass fraction in a pulsating white dwarf star,
with an internal precision of a few percent \cite{mnw00}. We used this
value to derive a preliminary measurement of the $^{12}{\rm
C}(\alpha,\gamma)^{16}{\rm O}$ reaction rate \cite{mwc01,msw02}, which
turned out higher than most published values \cite{kun01}. More work on
additional white dwarf stars and possible sources of systematic
uncertainty should help to resolve the discrepancy.

Our success with the parallel genetic algorithm leads us to believe that
many other problems of interest in astronomy and physics could benefit
from this approach. For models that can run in less than a few minutes on
currently available processors, and where automated execution is possible,
the parallel version of {\tt PIKAIA} can provide an objective and
efficient alternative to large grid searches without sacrificing the
global nature of the solution.  Although the number of model evaluations
required is still large compared to what can be accomplished in reasonable
wallclock time on a single desktop computer, Linux clusters are fast,
inexpensive, and are quickly becoming ubiquitous. When combined with
software like {\tt PIKAIA} that can exploit the full potential of such
distributed architectures, a new realm of modeling possibilities opens up.

\begin{acknowledge}
We are grateful to the High Altitude Observatory Visiting Scientist
Program for fostering this collaboration in a very productive environment
for two months during the summer of 2000. This work was supported by grant
NAG5-9321 from the Applied Information Systems Research Program of the
National Aeronautics \& Space Administration, and in part by the Danish
National Research Foundation through its establishment of the Theoretical
Astrophysics Center. The National Center for Atmospheric Research is
sponsored by the National Science Foundation.
\end{acknowledge}

\end{document}